\documentclass[a4paper,titlepage,11pt,english]{article}
\usepackage[utf8]{inputenc}
\usepackage[T1]{fontenc}
\usepackage{geometry}
\usepackage[english]{babel}
\usepackage{gensymb}
\usepackage{graphicx}
\usepackage{cite}
\usepackage{authblk}

\begin{document}

\title{Imaging light scattered by a subwavelength nanofiber, from near field to far field}

\author[$\dagger$]{Vivien Loo}
\author[$\dagger$]{Guillaume Blanquer}
\author[$\ddag$]{Maxime Joos}
\author[$\ddag$]{Quentin Glorieux}
\author[$\dagger$]{Yannick De Wilde}
\author[$\dagger$]{Valentina Krachmalnicoff\footnote{valentina.krachmalnicoff@espci.fr}}

\affil[$\dagger$]{Institut Langevin, ESPCI Paris, PSL Research University, CNRS, 1 rue Jussieu, F-75005, Paris, France}
\affil[$\ddag$]{Laboratoire Kastler Brossel, Sorbonne Universit\'{e}, CNRS, ENS-PSL Research University, Coll\`{e}ge de France, Paris 75005, France}

\maketitle

\begin{abstract}
We present a direct experimental investigation of the optical field distribution around a suspended tapered optical nanofiber by means of a fluorescent scanning probe. Using a 100~nm diameter fluorescent bead as a probe of the field intensity, we study interferences made by a nanofiber (400~nm diameter) scattering a plane wave (568~nm wavelength). Our scanning fluorescence near-field microscope maps the optical field over 36~$\mu$m$^2$, with $\lambda / 5$ resolution, from contact with the surface of the nanofiber to a few micrometers away. Comparison between experiments and Mie scattering theory allows us to precisely determine the emitter-nanofiber distance and experimental drifts.
\end{abstract}
\nopagebreak 

\section{Introduction}
An optical fiber tapered until its diameter is shorter than the wavelength will guide the light in such a way that a significant part of the electromagnetic field spreads outside the surface of the fiber\cite{tong2003,tong2004}. These modes are both propagating and evanescent, and the non-tapered ends of fibers are readily integrated into optical systems. They have been attracting much interest since the 2000s, for biosensors\cite{leung2007}, quantum information\cite{vetsch2010,sayrin2015}, coupling light to bigger structures (microspheres\cite{knight1997}, photonic crystals\cite{srinivasan2004}, microdisks\cite{kuo2014},...) to quantum emitters (cold atoms\cite{nayak2008,corzo2016}, CdSe quantum dots\cite{yalla2012}, nitrogen vacancy in diamond\cite{Liebermeister2014}, hBN flakes\cite{schell2017}, single molecules\cite{skoff2018}, ...). Nanofibers can also be processed\cite{nayak2011,nayak2013,li2017}, for instance into nanofiber cavities to enhance the coupling efficiency of an emitter on the surface into its fundamental mode\cite{schell2015}. More than just light wires, nanofibers confine the transverse electromagnetic field to the point that a longitudinal component appears\cite{lekien2004}, bonding the local polarization of light to its propagation direction; also known as spin-orbit coupling of light\cite{petersen2014} or spin-momentum locking\cite{mechelen2016}, this effect opens new possibilites in photonics\cite{Lodahl2017,joos2018}.

The most intuitive way of studying emitters on nanofibers is obviously to excite them through the fundamental guided mode and collect their near-field fluorescence through the same fiber, or their far-field emission in free space. But experimentally, a luminescence background is generated by the optical fiber itself since they are not pure silica (GeO$_2$ doped core, fluorine doped cladding, ...), and it often spectrally overlaps with the emitter signal. To get rid of this self-fluorescence background, the emitters are often excited sideways, from free space. Under such perpendicular illumination, the nanofiber scatters the incoming light and the excitation field on its surface is the result of interferences. It is crucial to know the spatial distribution of this field if one wants to optimize the emitter excitation.

In this Letter, we report direct observation of these interferences using a scanning fluorescence microscope\cite{krachmalnicoff2013,cao2015,Bouchet2016}. We probe the intensity of the optical field over a region of $16\times2.3~\mu m^2$ with deep sub-wavelength resolution. We compared experimental results to the case of a plane wave scattered by a nanofiber using Mie scattering theory, that provides semi-analytical solution \cite{barber1990} to this problem. Very good agreement between theoretical and experimental data validates our approach for the characterization of the optical field around suspended nanofibers. Moreover, we overcome the experimental challenge to pinpoint a nanoprobe with respect to a suspended nanofiber, leading to $\lambda/5$ accuracy from near-field to far-field.

\section{Experiment}

\begin{figure}[hb]
\begin{center}
\includegraphics[width=8cm]{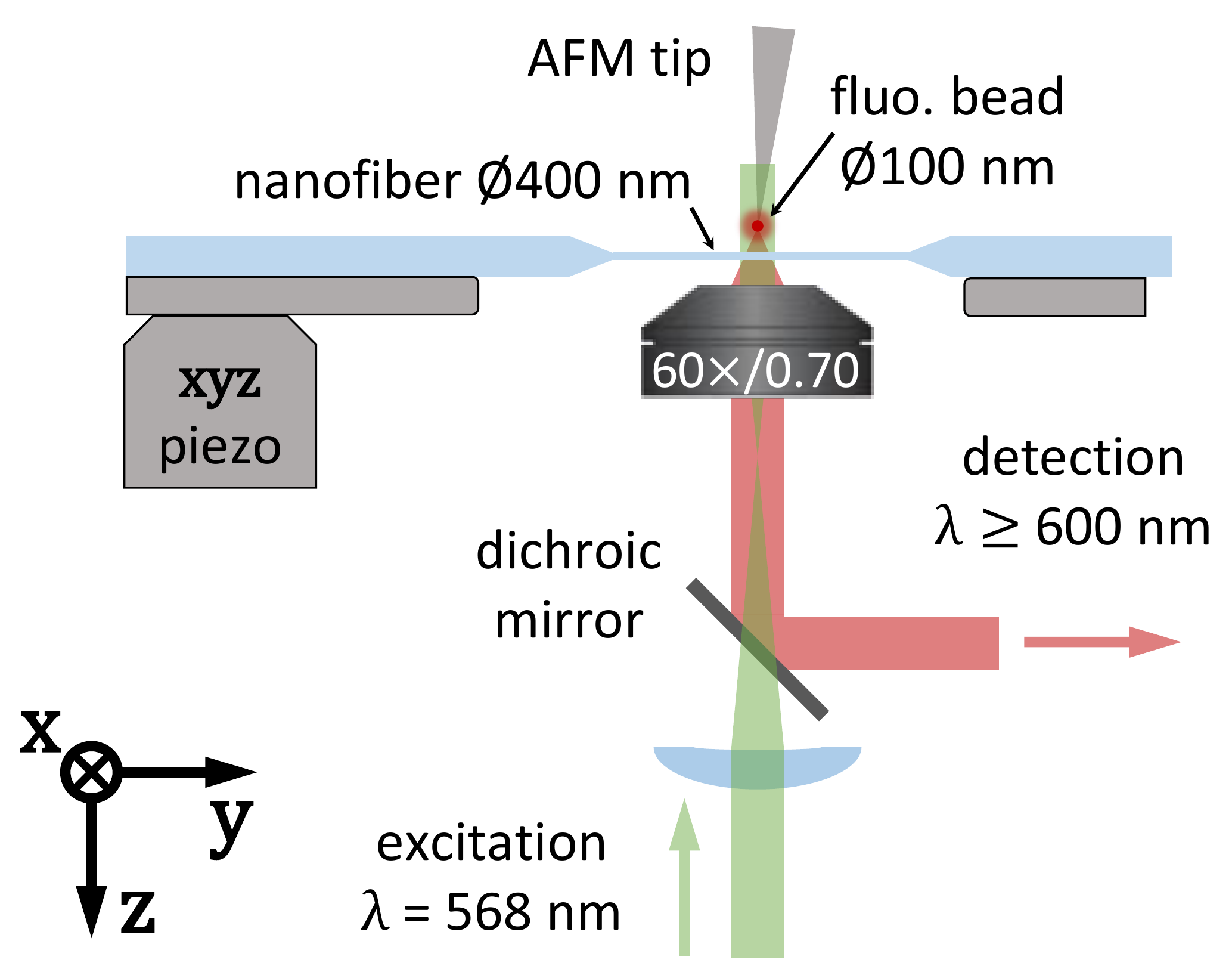}
\caption{Scheme of our fluorescence scanning probe apparatus. The confocal microscope is focused on a fluorescent bead grafted on the AFM tip. Scanning is done by moving the nanofiber with closed-loop piezoelectric actuators.}\label{fig1}
\end{center}
\end{figure}

The experimental setup is described in fig. \ref{fig1}. A linearly polarized laser ($\lambda = 568$~nm) propagating along the $z$-axis is focused on the back focal plane of a long working distance objective (numerical aperture NA~=~0.70). This produces a collimated beam characterized by a gaussian profile with a full width at half maximum of about 35~$\mu$m. 
The beam is scattered by a nanofiber ($d = 400$~nm, fabricated following reference\cite{hoffman2014}) suspended in the air along the $y$-axis. After interaction with the nanofiber, the intensity of the electromagnetic field exhibits fringes generated by the interference between the unscattered and the scattered beam. 

In order to directly probe the interference pattern, a fluorescent scanning near-field probe is set above the nanofiber. It consists of a 100~nm dye-doped polystyrene bead grafted on the tungsten tip of an atomic force microscope (AFM) \cite{krachmalnicoff2013}. The fluorescent bead absorbs the incident light and emits a fluorescence signal proportional to the excitation field at its very location (we work well below saturation). Collecting through the same objective used for the excitation, we acquire this emission signal as a function of the nanofiber position relative to the fluorescent probe. The latter must remain immobile in the objective focal plane and is conjugated with a single-photon avalanche diode (PDM-R, Micro Photon Devices~\cite{gulinatti_new_2012}) combined with a time-correlated single photon counting system (HydraHarp400, PicoQuant). 

\begin{figure}[h]
\begin{center}
\includegraphics[width=13cm]{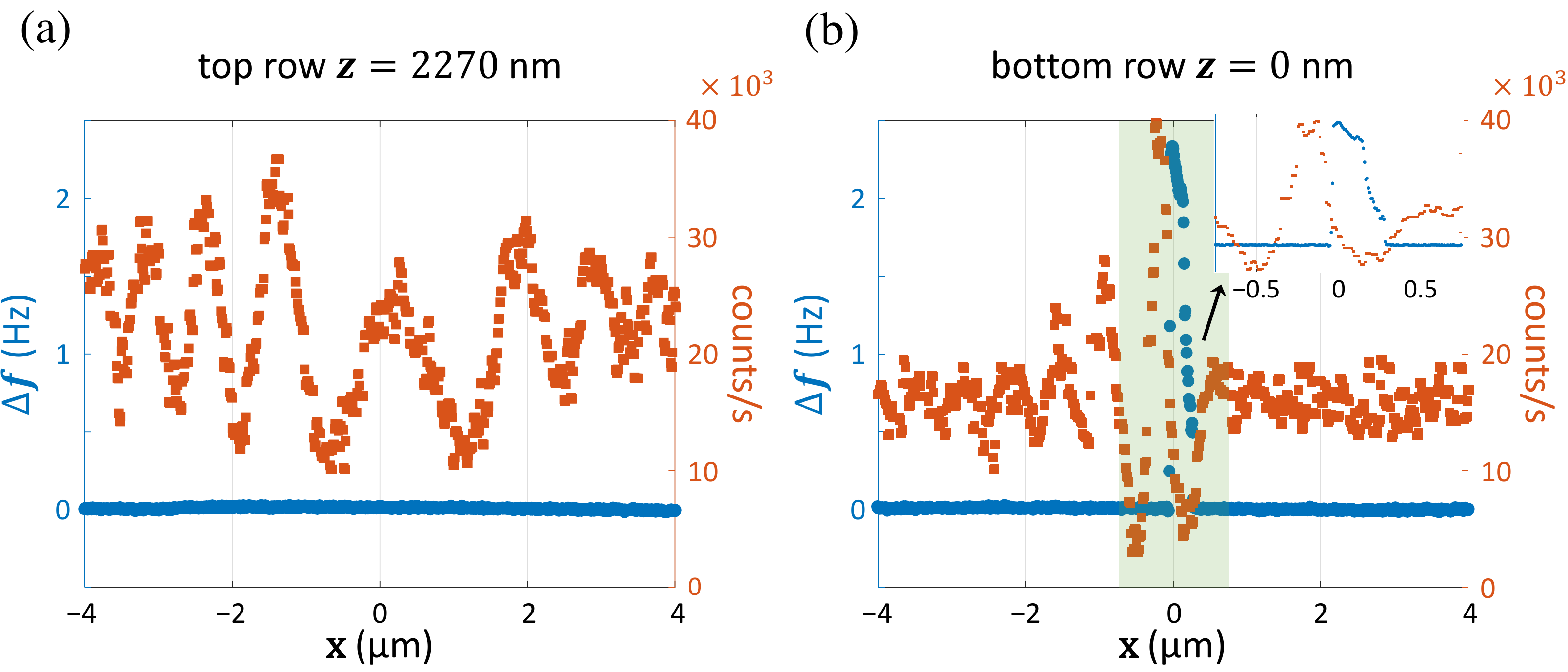}
\caption{Fluorescence signal in counts/s (red squares) and frequency shift $\Delta f$ in Hz (blue dots) along the $x$-axis, when (a) the tip is $z = 2770$~nm above the nanofiber, and (b) when we detect the first contact on the nanofiber at $z = 0$~nm. The inset highlights that the maximum of fluorescence does not coincide with the tip-fiber contact.}\label{fig2}
\end{center}
\end{figure}

Thanks to a $xyz$ piezoelectric scanner, the fiber repeatedly travels 16~$\mu$m back ($x\nearrow$) and forth ($x\searrow$) along $x$ by steps of 7.8~nm, then takes a 10~nm step closer to the fluorescent probe in the $z$-direction. Our AFM tip vibrates in the $xy$-plane, which coincides with the shear-force mode\cite{Karrai2000}. Starting from the largest distance between the nanofiber and the tip, we monitor its vibration frequency ($f \sim 32$ kHz) and proceed with the $xz$ scan of the fiber, pending contact between the AFM tip and the nanofiber, which is flagged by a typical shift $\Delta f \sim 2$ Hz. Figure \ref{fig2} shows the first row (a) and the last one (b) of such a scan. The first contact occurred during (b), as the nanofiber was moving backward($x\nearrow$). In these graphs, a red square on the scale of $x$-axis is about 100~nm, like the probe which is the limiting factor on spatial resolution.

Since the scan covers a vertical distance of a few microns, which is much larger than the distance at which shear-force interactions can be detected, the AFM is run without feedback. It means that we can detect when the AFM tip touches the fiber, but we do not have a reliable measurement of the probe-fiber distance at all times during the $xz$-scan. This distance will be precisely determined later on. For now, we simply set $z=0$ when the first tip-fiber contact occurs while scanning along $x$, and we scale the rest based on the calibration of the piezoelectric actuator.

Once the fluorescent bead is grafted on the tip, it is necessary to experimentally check its position with respect to the tip apex. This is done by comparing the intensity of the fluorescence signal and the tip vibration frequency close to contact. On the inset of fig. \ref{fig2}(b), one can measure a $\delta = 125$~nm gap between the maximum fluorescence intensity and the contact point. This is the horizontal deviation from the ideal graft, and it tells us on which side of the tip the fluorescent probe lies. This is of importance, since the tip is tilted (as depicted in fig. \ref{fig1}): it means that for $x<0$ the probe is directly exposed to the light scattered by the nanofiber, whereas for $x>0$ the tip end is between the nanofiber and the fluorescent bead.

Another important quantity to be characterized is the vertical distance of the grafting point from the extremity of the tip. Indeed, this distance will define the shorter emitter-nanofiber distance that will be accessible. As we will see later on, a comparison between experimental maps and Mie theory allows a precise determination of the shortest nanofiber-emitter distance achieved in this specific realization of the experiment.    

\begin{figure}[h]
\begin{center}
\includegraphics[width=13cm]{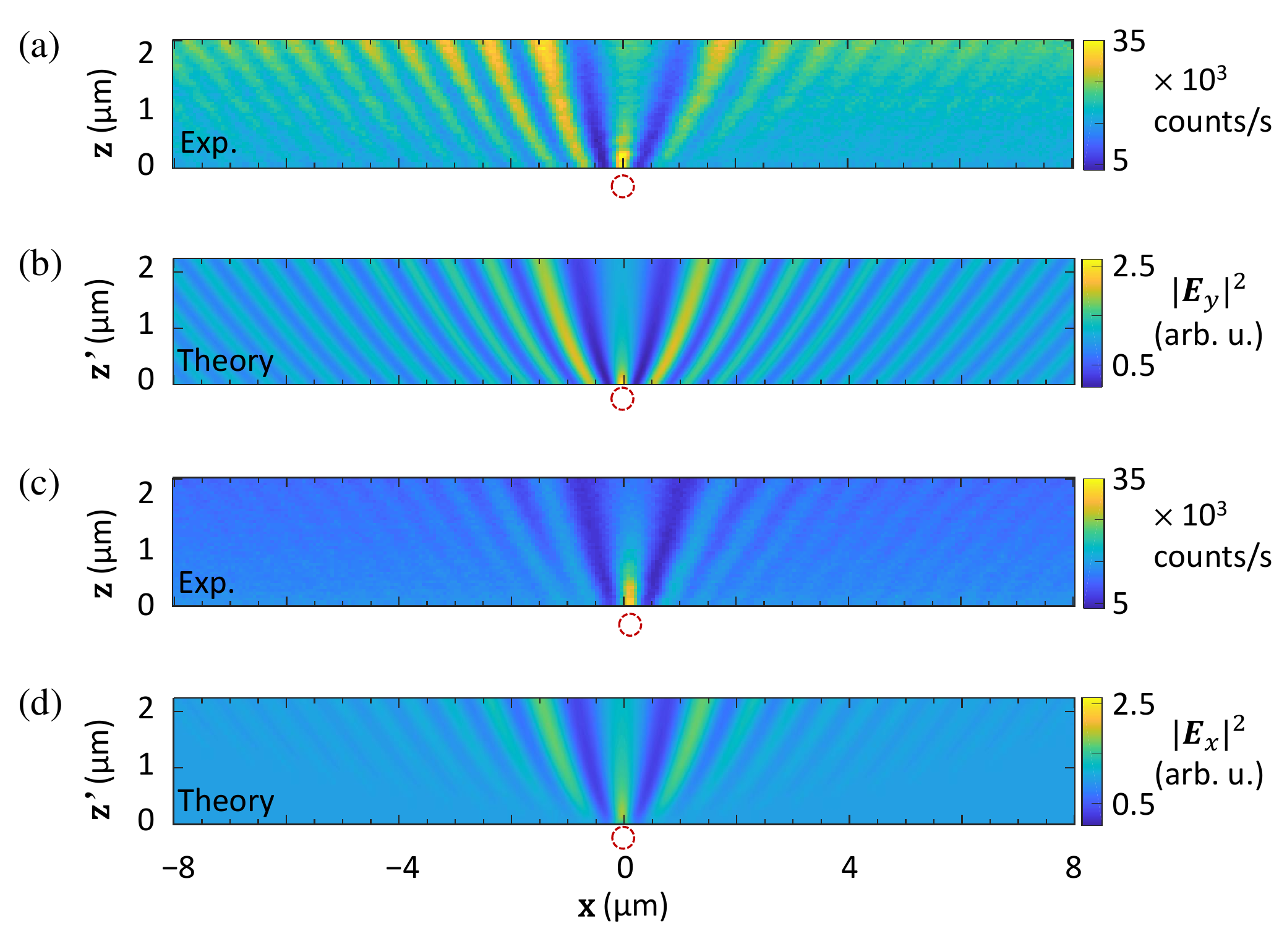}
\caption{Fluorescence signal over 36 $\mu$m$^2$ in the $xz$-plane; both axes have the same scale. The red dashed circles indicates the nanofiber position. In (a) the excitation beam is polarized parallel to the nanofiber (along $y$-axis), and in (c) perpendidular to it (along $x$-axis); pixels are 62~nm wide and 50~nm high ($8\times 5$ data binning). $z=0$ are set when the tip touches the nanofiber and $z$-axes are scaled using the actuator calibration. Using Mie theory, (b) and (d) show the calculated intensity of the electric field for the same polarizations respectively, this time $z'=0$ is at the surface of the nanofiber.}\label{fig3}
\end{center}
\end{figure}

Figure \ref{fig3} shows experimental (a,c) and theoretical (b,d) maps of the excitation field intensity in the $xz$-plane. Due to the geometry of the problem, we studied two cases: in fig. \ref{fig3}(a, b) the laser is polarized parallel to the fiber (along $y$-axis), and perpendicular to it (along $x$-axis) in fig. \ref{fig3}(c, d). Experimental maps are obtained by compiling the $x$-axis forward scan rows (like in fig. \ref{fig2}(a)), from  far-field ($z=4 \lambda = 2270$~nm) down to near-field until contact ($z=0$~nm).  Interferences are clearly visible and the agreement between theoretical and experimental maps is evident. Note that maps with different polarizations are plotted with the same intensity range to help with the comparison. 

While in the parallel polarized case fig. \ref{fig3}(a,b) fringes contrast and intensity are high, in the perpendicularly polarized case fig. \ref{fig3}(c,d), the contrast fades away as the scattering angle and the distance to the fiber increase. Indeed, scattering at large angles means propagating along $x$ after scattering on the fiber. This is unlikely in free space for scattering on a dielectric fiber when the incident polarization is linear along $x$ as well.

Note that fringes contrast and period in the experimental maps are slightly different on each side of the nanofiber. This can be explained by the asymmetry of probe position relative to the apex of the tip, as mentioned when discussing fig. \ref{fig2}(b).

The keen eye will also notice that around $x \sim 0$, the signal ripples along $z$ with a period close to $\lambda/2$; due to some possible interplay between the tip and the nanofiber forming a Fabry-Perot cavity. To back this up, on a different set of data (not shown here), we observed that these oscillations are more visible when the tip, nanofiber, and polarization are all parallel to each other.

To take into account the anisotropic scattering produced by the nanofiber, Mie theory is most suited, given the symmetry of the problem (cylindrical), the ratio of the diameter of the nanofiber with respect to wavelength ($d=0.70 \lambda$), and the distance range over which we want to know the optical field (0 to $4 \lambda$). Semi-analytical solutions\cite{barber1990} are shown in fig. \ref{fig3}(b) and (d) for parallel and perpendicular polarizations, respectively. In this model, an incident plane wave interferes with the optical field scattered by the transparent cylinder (i.e. the nanofiber), which results in a complex set of fringes.

Theoretical calculations need three fixed parameters: the wavelength is set $\lambda = 568$~nm; the fiber index is given by its reseller $n = 1.46$; and the diameter of the nanofiber was measured by scanning electron microscopy $d = 400 \pm 20$~nm. The only free parameter that we can adjust to match experimental and theoretical results is the actual emitter-fiber distance, the latter being not continuously measured during the scan with respect to a reference plane as done in standard scanning probe measurements. The experimental $z$-axis will therefore be corrected by a quantity $\Delta z$ that is expected to have two components.

Firstly, a constant one, which was mentioned before: the fluorescent probe is not ideally placed on the apex of the tip. Moreover, the tip cannot hover closer to a suspended nanofiber than a minimum distance below which the fiber suddenly snaps into the tip due to attractive forces between the tip and the fiber which is a soft spring. This situation is rather different to the typical case where the AFM scans on a rigid substrate such as a glass coverslip.

Secondly, a time dependent component: acquiring a 36~$\mu$m$^2$ scan takes sixty minutes, drifts are to be expected, be it from the actuators or from the nanofiber stretched in the air.
\begin{figure}[h]
\begin{center}
\includegraphics[width=13cm]{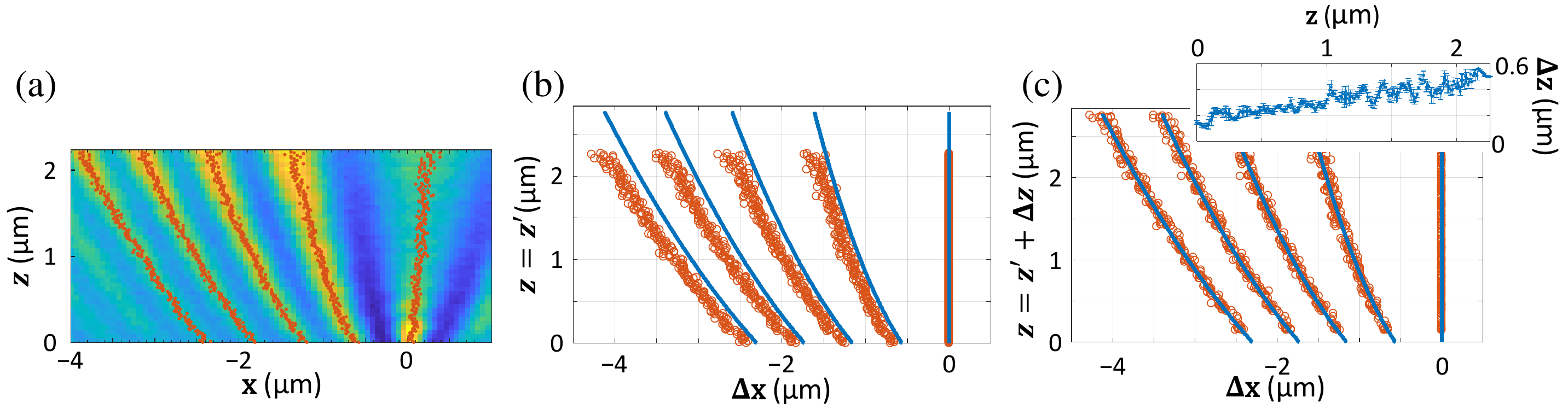}
\caption{(a) Part of figure \ref{fig3}(a) showing where we locate the maximum of several peaks (red dot) from fig. \ref{fig3}(a). These positions are plotted in (b) (red circles) together with the theoretical positions (blue solid lines), as a function of their relative distance $\Delta x$ to the central peak, assuming the experimental distance probe-nanofiber $z$ natively matches the theoretical axis $z'$. (c) Same graph as (b) plotted with the actual probe-fiber distance. The correction $\Delta z$ applied to $z$-axis is shown in inset.}\label{fig4}
\end{center}
\end{figure}

Figure \ref{fig4} illustrates how we determine the probe-fiber distance. We start by picking the positions of the first few local maxima of fig. \ref{fig3}(a) as illustrated on fig. \ref{fig4}(a), and we plot them (red circles) along their theoretical homologs (blue line) on fig. \ref{fig4}(b). Initially we do so assuming that our experimental $z$-axis perfectly matches its theoretical counterpart $z'$. Since the experimental $x$-axis is also arbitrary and subject to drift relative to the suspended nanofiber, the peaks positions are given in terms of $\Delta x$, their relative distance to the central maximum for a given height during the scan in the $xz$-plane. Figure \ref{fig4} (b) shows that all the peaks would be experimentally too far away from the center if the probe was at the apex of the tip, without any drift (case $z=z'$).

We proceed to numerically find the best offset $\Delta z = z - z'$ such as all of the four left peaks ($\Delta x<0$) match their theoretical positions. The result of this procedure is shown in the inset of fig. \ref{fig4}(c). $\Delta z$ is found for every row of the scan, with a moving average filter (over five $z$ steps, that is 50~nm). The almost linear variation of $\Delta z$ as a function of $z$ (which corresponds also to the scan time since the probe moves at constant speed) well confirms the hypothesis of a mechanical or thermal slow drift. 

We determine that the snap-in occurs at $\Delta z = 150~nm$, that being the distance between the fluorescent emitter and the nanofiber; the tip apex itself is most probably closer. If our system had no drift at all, $\Delta z$ would stay constant at this snap-in distance. Instead, we observe that the probe and fiber slowly drift closer to each other: about 400~nm over an hour. Fig. \ref{fig4}(c) shows the experimental peaks relative positions matched to the theory by correcting each $z$ by its corresponding $\Delta z$. 
For each row of the scan, a single value of $\Delta z$ makes all peaks match to theory at the same time. We reproduced this data analysis on similar experiments: $\Delta z$ at contact depends on each tip-probe graft, we reported the dataset with the shortest distance; drift was observed in every case.

\section{Conclusion}
We used a fluorescence scanning near-field microscope on a suspended nanofiber ($d=400$~nm) to investigate how it scatters a plane wave ($\lambda = 568$~nm). To our knowledge, direct mapping, in both the transverse and longitudinal directions, of a plane wave scattered by a subwavelength cylinder was never reported in the distance range we explored ($\lambda/4$ to $4 \lambda$), neither in the visible range \cite{warken2004,Little2014} nor in the microwave realm \cite{kozaki1982}. This illumination scheme allows to compare analytical solutions from Mie scattering theory to our experimental results. The former and the latter are very consistent, which confirms the technique is sound to probe complex nanofiber/emitter systems. Trapped atoms in the evanescent field of a nanofiber are often probed by a beam orthogonal to the nanofiber\cite{nayak2008} and experimentally investigating the resulting inhomogeneous near-field at the atoms location is useful.

In the realization of the experiment presented here, we could approach the fluorescent emitter and the nanofiber down to 150~nm before making contact. AFM feedback signal proves that the fluorescent emitter was grafted far from the extremity of the tip, meaning that the effective snap-in distance (i.e. the distance at which the extremity of the tip touches the nanofiber) is shorter. By optimizing the grafting procedure, tip shape and material, we could reach smaller distances (tens of nm) between the fluorescent emitter and the extremity of the tip. This is encouraging since it proves that the fluorescent probe can accurately scan the nanofiber fundamental guided mode, whose evanescent field spreads 200~nm outside the surface, for the diameter and wavelength studied in the present work. By the same token, our technique can be used to map the modification of the guided electromagnetic field induced by the presence of a nano-object (such as for example a metallic nanorod) on the nanofiber.

Our setup has lifetime measurement capabilities, and the local density of states (LDOS), which is inversely proportional to the lifetime of a fluorescent emitter \cite{Carminati2015}, can be mapped with a resolution of few tens of nanometers while the probe is in the near field of a nanostructured environment \cite{krachmalnicoff2013,Bouchet2016}. This measurement gives access to a direct visualization of light-matter interaction at the nanometer scale and can be performed by grafting to the tip fluorescent beads or single photon emitters, such as for example a single quantum dot or a single NV center in a diamond nanocrystal. By mapping the LDOS simultaneously with the fluorescence intensity, a quantitative measurement of the coupling rate of fluorescence photons to the nanofiber guided mode and its comparison to other fluorescence decay channels \cite{cao2015} will soon become within reach. 

\section*{Funding}
PSL Research University in the framework of the project COSINE; LABEX WIFI (Laboratory of Excellence ANR-10-LABX-24) within the French Program Investments for the Future under reference ANR-10- IDEX-0001-02 PSL*; Programme Emergences 2015 of the City of Paris.

\section*{Acknowledgments} The authors thank I. Rech, A. Gulinatti and A. Giudice for providing the PMD-R (Micro Photon Devices) detector. 

\bibliographystyle{unsrt}
\bibliography{biblio}
\end{document}